\documentclass[submission,copyright,creativecommons]{eptcs}
\providecommand{\event}{QPL 2011} 
\usepackage{amsfonts}
\usepackage{amssymb}
\usepackage{amsmath}

\newtheorem{definition}{Definition}[section]
\newtheorem{remark}{Remark}[section]
\newtheorem{example}{Example}[section]
\newtheorem{theorem}{Theorem}[section]
\newtheorem{proposition}[theorem]{Proposition}
\newtheorem{lemma}[theorem]{Lemma}
\newtheorem{corollary}[theorem]{Corollary}

\allowdisplaybreaks[4]
\title{{Turing machines based on unsharp quantum logic}
\thanks{
  This work was supported by an NSFC project under Grant No.\ 61073023,
  by a 973-Project under Grant No.\ 2009CB320701, by the K.~C.~Wong
  Education Foundation, Hong Kong, and by the National Center for
  Mathematics and Interdisciplinary Sciences, CAS.}}
\author{Yun Shang
\institute{Institute of Mathematics,\\ AMSS, CAS\\ Beijing, P.R.China}
\email{shangyun602@163.com}
\and
Xian Lu
\institute{Institute of Mathematics,\\ AMSS, CAS\\ Beijing, P.R.China}
\email{luxian@amss.ac.cn}
\and
Ruqian Lu
\institute{Institute of Mathematics,\\ AMSS, CAS\\ Beijing, P.R.China}
\institute{CAS Key Lab of IIP\\
Institute of Computing Technology, CAS.\\
Beijing, P.R.China}
\email{rqlu@math.ac.cn}
}
\def\titlerunning{Turing machines based on unsharp quantum logic}
\def\authorrunning{Y. Shang, X. Lu\& R.Q. Lu}
\begin{document}
\maketitle

\begin{abstract}
In this paper, we consider Turing machines based on unsharp quantum logic. For a lattice-ordered quantum multiple-valued (MV) algebra $\mathcal{E}$, we introduce $\mathcal{E}$-valued non-deterministic Turing machines ($\mathcal{E}$NTMs) and $\mathcal{E}$-valued deterministic Turing machines ($\mathcal{E}$DTMs). We discuss different $\mathcal{E}$-valued recursively enumerable languages from width-first and depth-first recognition. We find that width-first recognition is equal to or less than depth-first recognition in general. The equivalence requires an underlying $\mathcal{E}$ value lattice to degenerate into an MV algebra. We also study variants of $\mathcal{E}$NTMs. $\mathcal{E}$NTMs with a classical initial state and $\mathcal{E}$NTMs with a classical final state have the same power as $\mathcal{E}$NTMs with quantum initial and final states. In particular, the latter can be simulated by $\mathcal{E}$NTMs with classical transitions under a certain condition. Using these findings, we prove that $\mathcal{E}$NTMs are not equivalent to $\mathcal{E}$DTMs and that $\mathcal{E}$NTMs are more powerful than $\mathcal{E}$DTMs. This is a notable difference from the classical Turing machines.
\end{abstract}

\section{Introduction}

In traditional von Neumann quantum logic, $\mathcal{P(H)}$ (the set of all projection operators of a Hilbert space $\mathcal H$) is regarded as a set of quantum events. It constitutes an orthomodular lattice, which is the main algebraic model in quantum logic. However, since the set of projection operators is not the maximal set of possible events according to the statistical rules of open quantum systems, $\mathcal{E(H)}$ (the set of all positive operators dominated by the identity on $\mathcal H$) becomes a new quantum event set. Since any event in ${\mathcal{P(H)}}$ always satisfies the non-contradiction law, traditional quantum logic is called sharp quantum logic. Quantum events represented by $\mathcal{E(H)}$ do not satisfy the non-contradiction law, and the quantum logic corresponding to $\mathcal{E(H)}$ is called unsharp quantum logic. Many algebraic structures have been proposed to characterize unsharp quantum events, and effect algebras \cite{FB94} are the main model for unsharp quantum logic. Multiple-valued (MV) algebras, as algebraic models of multiple-valued logic, play an analogous role to that of Boolean algebras in sharp quantum logic \cite{DP00}. Quantum MV (QMV) algebras are another important type of unsharp quantum structure \cite{G96}.

For abstract mathematical machines, automata theory is one of the main branches in classical computing theory. It mainly consists of finite-state automata, pushdown automata, and Turing machines. Although classical computing theory can be regarded as part of classical mathematical theory, the logical foundation of automata theory is still Boolean logic. Quantum logic differs from classical logic and quantum devices should obey their own logic. Hence, an interesting question arises: can we set up a quantum computation theory based on quantum logic? Ying et al. set up finite-state automata and pushdown automata theories based on sharp quantum logic \cite{Y07,Q04}. They found that some important properties similar to classical automata are universally valid if and only if the underlying truth value lattice degenerates to a Boolean algebra. Li proved that deterministic finite automata and non-deterministic finite automata based on sharp quantum logic are equivalent, independent of the distributive law \cite{Li10}. Since unsharp quantum logic is more universal than sharp quantum logic, Shang et al. set up finite-state automata and pushdown automata theories based on unsharp quantum logic. They found that some important properties similar to classical automata are universally valid if and only if the underlying truth value lattice degenerates to an MV algebra \cite{S09,S11}.

Since Turing machines are a core concept in the study of computing theory, we continue to study Turing machines based on unsharp quantum logic. Deutsch proposed quantum Turing machines from a quantum mechanics point of view \cite{D85} and Perdrix generalized this to observable quantum Turing machines \cite{P07}. Perdrix and Jorrand introduced classically controlled Turing machines \cite{P11}. Bernstein et al. addressed universal quantum Turing machines \cite{B97}. However, the logical foundation for these machines is still Boolean logic. The relation between the above Turing machines and the proposed Turing machines is similar to the relation between quantum mechanics and quantum logic.

In this paper, we mainly consider two algebraic models of unsharp quantum logic for Turing machines, namely extended lattice-ordered-effect algebras and lattice-ordered QMV algebras. Here we call them $\mathcal{E}$-valued lattices. Although similar to finite-state automata and pushdown automata based on unsharp quantum logic, some important properties of Turing machines based on unsharp quantum logic depend heavily on the distributivity of the underlying logic. However, we find that $\mathcal{E}$-valued non-deterministic Turing machines ($\mathcal{E}$NTMs) are not equivalent to $\mathcal{E}$-valued deterministic Turing machines ($\mathcal{E}$DTMs) even if the distributivity of the underlying logic holds. This is a characteristic difference from classical Turing machines.

The remainder of the paper is organized as follows. Section \ref{sec:2} provides some algebraic results used later in the paper. In Section \ref{sec:3}, we introduce the concepts of $\mathcal{E}$NTMs and $\mathcal{E}$DTMs. We also define two patterns of recursively enumerable language recognition for unsharp quantum Turing machines: width-first (namely, parallel) recognition and depth-first (namely, sequential) recognition, similar to the case in unsharp quantum automata. We prove that the width-first recognizability of a recursively enumerable quantum language is always equal to or less than its depth-first recognizability. We find that equivalence requires the underlying $\mathcal{E}$ value lattice to degenerate to an MV algebra. In section \ref{sec:4}, we discuss variants of unsharp quantum Turing machines. $\mathcal{E}$NTMs with a classical initial state and $\mathcal{E}$NTMs with a classical final state have the same power as $\mathcal{E}$NTMs with quantum initial and final states. In particular, under a certain condition, the latter can be simulated by an $\mathcal{E}$NTM with classical transitions. Using these results, we find that $\mathcal{E}$NTMs are more powerful than $\mathcal{E}$DTMs. This is different from the result in classical computing theory. Section \ref{sec:5} presents our main conclusion.


\section{Extended lattice-ordered-effect algebras and lattice-ordered QMV algebras}\label{sec:2}

First, we provide some notions and results in unsharp quantum logic.

\begin{definition}\rm\cite{DP00}
A supplement algebra (S-algebra for short) is an algebraic structure $\mathcal{E}=(E,\boxplus,',\mathbf{0},\mathbf{1})$ consisting of set $M$ with two constant elements $\mathbf{0},\mathbf{1}$, a unary operation $'$ and a binary operation $\boxplus$ on $M$ satisfying the following axioms:

\noindent\begin{tabular}{@{}llll}
(S1) & $a\boxplus b=b\boxplus a$. & (S2) & $a\boxplus(b\boxplus c)=(a\boxplus b)\boxplus c$.\\
(S3) & $a\boxplus a'=\mathbf{1}$. & (S4) & $a\boxplus\mathbf{0}=a$.\\
(S5) & $a''=a$. & (S6) & $a\boxplus \mathbf{1}=\mathbf{1}$.
\end{tabular}
\end{definition}

An MV algebra is an S-algebra that satisfies:

\smallskip\noindent
(MV) $(a'\boxplus b)'\boxplus b=(a\boxplus b')'\boxplus a$.
\smallskip

For an S-algebra, we define the following three binary operations: $a\odot b=(a'\boxplus b')'$, $a\sqcap b=(a\boxplus b')\odot b$, and $a\sqcup b=(a\odot b')\boxplus b$.

A QMV algebra is an S-algebra that satisfies:

\smallskip\noindent
(QMV1) $a\sqcup(b\sqcap a)=a$.\\
(QMV2) $(a \sqcap b)\sqcap c=(a\sqcap b)\sqcap (b\sqcap c)$.\\
(QMV3) $a\boxplus [b\sqcap (a\boxplus c)']=(a\boxplus b)\sqcap(a\boxplus (a\boxplus c)']$.\\
(QMV4) $a\boxplus (a'\sqcap b)=a\boxplus b$.\\
(QMV5) $(a'\boxplus b)\sqcup (b'\boxplus a)=\mathbf{1}$.
\smallskip 

A partial relation $\leq$ in QMV algebra can be defined as $a\leq b$ iff $a=a\sqcap b$.

It is clear that a QMV algebra is not necessarily a lattice under the operations $\sqcap$ and $\sqcup$. If $\mathcal{E}$ forms a lattice with $\leq$, it is called a lattice-ordered QMV algebra, where $\wedge$ denotes the infimum operation and $\vee$ denotes the supremum operation in the lattice. A QMV algebra $M$ is quasilinear if $a\not\leq b$ implies $a\sqcap b=b$. A QMV algebra (or an MV algebra) $M$ is linear if $\forall a,b\in M$, either $a\leq b$ or $b\leq a$. There exists a QMV algebra that is not quasilinear (Example 1, \cite{G95}). Every MV algebra is a QMV algebra; however, there exists a QMV algebra that is not an MV algebra (Example 2.7, \cite{S09}).

An effect algebra is a set $P$ with two particular elements $0,1$ $(0\neq1)$ and with a partial binary operation $\oplus: P\times\,P\longrightarrow\,P$ such that, for all $a,b,c\in\,P$:

\smallskip
\noindent(E1) If $a\oplus\,b\in\,P$, then $b\oplus\,a\in\,P$ and $a\oplus\,b=b\oplus\,a$.\\
(E2) If $b\oplus\,c\in\,P$ and $a\oplus(b\oplus\,c)\in\,P$, then $a\oplus\,b\in\,P$ and $(a\oplus\,b)\oplus\,c\in\,P$ and $a\oplus(b\oplus\,c)=(a\oplus\,b)\oplus\,c$.\\
(E3) For any $a\in\,P$ there is a unique $b\in\,P$ such that $a\oplus\,b$ is defined and $a\oplus\,b=1$.\\
(E4) If $1\oplus\,a$ is defined, then $a=0$.

\begin{example}\rm
Let $\varphi=( E,\oplus,0,1)$ be an effect algebra. The operation $\oplus$ can be extended to a total operation $\boxplus:E\times E\longrightarrow E$ by defining
$$a\boxplus b=\left\{
\begin{array}{ll}
a\oplus b,&\;\hbox{if}\;(a\oplus b)\, \hbox{is defined}\\
1,&\;\hbox{otherwise.}
\end{array}\right.$$
We denote the resulting structure by $\bar{\varphi}=(E,0,1,\boxplus)$ and call it an extended-effect algebra. It is easy to see that an extended-effect algebra $\bar{\varphi}$ preserves the order of the effect algebra and is equivalent to a quasilinear QMV algebra \cite{G95}.
\end{example}

\begin{theorem}\rm\cite{S09}\label{thm:3}
Let $\mathcal{E}=(E,\boxplus,',\mathbf{0},\mathbf{1})$ be a
lattice-ordered QMV algebra. The following conditions are equivalent:

\begin{enumerate}
\item[(i)] $\mathcal{E}$ is an MV algebra.
\item[(ii)] $(a\boxplus b)\wedge(a\boxplus c)=a\boxplus(b\wedge c)$ for any $a,b,c\in E$.
\end{enumerate}
\end{theorem}

\begin{theorem}\rm\cite{S09}\label{thm:4}
Let $\mathcal{E}=(E,\boxplus,',\mathbf{0},\mathbf{1})$ be an extended
lattice-ordered-effect algebra. The following conditions are
equivalent:

\begin{enumerate}
\item[(i)] $\mathcal{E}$ is a linear MV algebra.
\item[(ii)] $(a\boxplus b)\wedge(a\boxplus c)=a\boxplus(b\wedge c)$ for any $a,b,c\in E$.
\end{enumerate}
\end{theorem}


\section{Unsharp quantum Turing machines}\label{sec:3}
If we let unsharp quantum logic denote the truth value set of the propositions, we can set up Turing machines based on unsharp quantum logic. In the following, $\mathcal{E}$ denotes a lattice-ordered QMV algebra. If $\mathcal{E}$ denotes an extended lattice-ordered-effect algebra, we can obtain Turing machines based on an extended lattice-ordered-effect algebra without changing anything.

\begin{definition}\label{def:1}\rm
An $\mathcal{E}$-valued non-deterministic Turing machine ($\mathcal{E}$NTM) is a septuple:
$M=(Q,\Sigma,\Gamma,\delta,B,\\ I,T)$, where

\begin{enumerate}
\item[1.] $Q$ is a finite nonempty-state set.
\item[2.] $\Sigma$ is the finite set of input symbols.
\item[3.] $\Gamma$ is the complete set of tape symbols; $\Sigma\subseteq\Gamma/B $.
\item[4.] $\delta:Q\times\Gamma\times Q\times\Gamma\times\{L,S,R\}\longrightarrow\mathcal{E}$ is the transition function. The symbols $L$, $R$ and $S$ indicate that the head of the $\mathcal{E}$NTM moves left or right or remains stationary, respectively.
\item[5.] $B$ is the blank symbol. The blank symbol appears initially in all but the finite number of initial cells that hold input symbols.
\item[6.] $I:Q\longrightarrow\mathcal{E}$ is the initial-state function.
\item[7.] $T:Q\longrightarrow\mathcal{E}$ is the final- or accepting-state function.
\end{enumerate}
\end{definition}
As defined for classical Turing machines, a configuration or instantaneous description (ID) of an $\mathcal{E}$NTM $M$ is a sequence $C=\alpha_1 q\alpha_2$, where $q\in Q$ and $\alpha_1\alpha_2$ is the finite sequence between the leftmost and rightmost nonblank symbols. We denote the state of $C$ by $St(C)$ and denote $ID(M)$ as the set of all instantaneous descriptions of $M$. An $\mathcal{E}$NTM in ID $\alpha_1 q\alpha_2$ means the current state is $q$ and the reading head is looking at the first symbol of $\alpha_2$. The value of $M$ transforming from $C_1$ to $C_2$ is described as
$$ \delta^\star(C_1,C_2)=
\begin{cases}
\delta(p,a,q,b,L),& \mbox{if }C_1=\alpha cpa\beta\mbox{ and }C_2=\alpha qcb\beta\\
\delta(p,a,q,b,S),& \mbox{if }C_1=\alpha pa\beta\mbox{ and }C_2=\alpha qb\beta\\
\delta(p,a,q,b,R),& \mbox{if }C_1=\alpha pa\beta\mbox{ and }C_2=\alpha bq\beta\\
1,& \mbox{otherwise,}
\end{cases}
$$
where $a,b,c\in\Gamma$ and $\alpha,\beta\in\Gamma^*$ such that the leftmost symbol of $\alpha$ and the rightmost symbol of $\beta$ are not $B$. $\vdash(C_1,C_2)=(p,a,q,b,D)$ denotes that the $\mathcal{E}$NTM can transform $C_1$ to $C_2$ through the transition $(p,a,q,b,D)$.

Similar to finite-state automata theory based on unsharp quantum logic, by interacting $\wedge$ and $\boxplus$, we can adapt depth-first and width-first methods for defining the degree of acceptance of languages recognized by Turing machines. In fact, these correspond to parallel recognition and sequential recognition. We prove that the methods coincide only when the truth lattice is an MV algebra.

\begin{definition}\rm
A path of an $\mathcal{E}$NTM $M$ is a finite sequence of IDs.
\end{definition}

\begin{definition}\label{def:2}\rm
The $\mathcal{E}$-valued language accepted by an $\mathcal{E}$NTM $M$ in a depth-first manner is defined as:
\begin{equation}\label{eqn:1}
\begin{split}
|M|_d(s)=&\bigwedge_{n\ge1}\bigwedge_{C_i}\bigwedge_{q_0\in Q}I(q_0)\boxplus
\delta^\star(q_0s,C_1)\boxplus\delta^\star(C_1,C_2)\boxplus\cdots
\boxplus T(St(C_n))
\end{split}
\end{equation}
for any $s\in\Sigma^+$.
\end{definition}

\begin{definition}\label{def:3}\rm
The $\mathcal{E}$-valued language accepted by an $\mathcal{E}$NTM $M$ in a width-first way is defined as:
\begin{equation}\label{eqn:2}
\begin{split}
|M|_w(s)=&\bigwedge_{n\ge1}\bigg[\bigwedge_{C_n}\bigg(\cdots\bigg(\bigwedge_{C_2}\Bigg(\bigwedge_{C_1}
\bigg(\bigwedge_{q_0}I(q_0)\boxplus\delta^\star(q_0s,C_1)\bigg)\boxplus\delta^\star(C_1,C_2)\bigg)\\
&\boxplus\delta^\star(C_2,C_3)\bigg)\cdots\bigg)\boxplus T(St(C_n))\bigg]
\end{split}
\end{equation}
for any $s\in\Sigma^+$.
\end{definition}

\begin{remark}\rm\label{rmk:1}
Similar to classical Turing machines, an $\mathcal{E}$NTM $M$ halts when it reaches some state $q$ with $T(q)<1$ or obtains some ID $C$ with $T(St(C))=1$ and $\delta^\star(C,C')=1$ for any ID $C'$. Each path in Equations (\ref{eqn:1}) and (\ref{eqn:2}) is required to halt. If the machine does not halt for some input $s$ in all paths, then the $\mathcal{E}$-value of $s$ accepted by $M$ is not defined.
\end{remark}

\begin{definition}\rm
An $\mathcal{E}$-valued deterministic Turing machine ($\mathcal{E}$DTM) is an $\mathcal{E}$NTM whose transition function $\delta$ satisfies the following: for any $p\in Q$ and $a\in\Gamma$, there exists at most one set $\{q,b,D\}$ such that $\delta(p,a,q,b,D)\ne 1$.
\end{definition}

The classes of all $\mathcal{E}$NTMs and $\mathcal{E}$DTMs over alphabet $\Sigma$ are denoted by NTM$(\mathcal{E},\Sigma)$ and DTM$(\mathcal{E},\Sigma)$, respectively. We denote $L_d^T(\mathcal{E},\Sigma)=\{|M|_d:M\in\mbox{NTM}(\mathcal{E},\Sigma)\}$ and $L_w^T(\mathcal{E},\Sigma)=\{|M|_w:M\in\mbox{NTM}(\mathcal{E},\Sigma)\}$.

\begin{definition}\rm
A partial function $L:\Sigma^+\rightarrow\mathcal{E}$ is called an $\mathcal{E}$-valued d-recursively enumerable (d-RE) language or an $\mathcal{E}$-valued w-recursively enumerable (w-RE) language if $L\in L_d^T(\mathcal{E},\Sigma)$ or $L\in L_w^T(\mathcal{E},\Sigma)$, respectively.
\end{definition}

\begin{proposition}\rm
\begin{enumerate}
\item[(i)] $|M|_w\le |M|_d$ for any $\mathcal{E}$NTM $M$.
\item[(ii)] $|M|_w=|M|_d$ for any $\mathcal{E}$NTM $M$ iff $\mathcal{E}$ is an MV algebra.
\end{enumerate}
\end{proposition}
{\bf Proof :} Point (i) is obvious since $a\boxplus(b\wedge c)\leq(a\boxplus b)\wedge(a\boxplus c)$ for $a,b,c\in\mathcal{E}$ in general. (ii) If $\mathcal{E}$ is an MV algebra, then $\boxplus$ distributes over $\wedge$, so $|M|_w=|M|_d$. Conversely, for any $a,b,c\in\mathcal{E}$ we construct an $\mathcal{E}$NTM $M=(\{q_0,q_1,q_2\},\Sigma,\Gamma,\delta,B,I,T)$ as follows. For some $\sigma\in\Sigma$,
\begin{align*}
&I(q_0)=b,I(q_1)=c,I(q_2)=1,\;T(q_0)=1,T(q_1)=1,T(q_2)=a\\
&\delta(q_0,\sigma,q_2,\sigma,R)=\delta(q_1,\sigma,q_2,\sigma,R)=0
\end{align*}
and $\delta=1$ for the rest. For the input $s=\sigma$, all the effective paths are $q_0\sigma\vdash\sigma q_2$ and $q_1\sigma\vdash\sigma q_2$. Thus, $|M|_d(\sigma)=(I(q_0)\boxplus\delta^\star(q_0\sigma,\sigma q_2)\boxplus T(q_2))\wedge(I(q_1)\boxplus\delta^\star(q_1\sigma,\sigma q_2)\boxplus T(q_2))=(a\boxplus b)\wedge(a\boxplus c)$. From the definition it is easy to see that $|M|_w(\sigma)=[(I(q_0)\boxplus\delta^\star(q_0\sigma,\sigma q_2))\wedge(I(q_1)\boxplus \delta^\star(q_1\sigma,\sigma q_2))\boxplus T(q_2)]=(b\wedge c)\boxplus a$. Therefore, $a\boxplus (b\wedge c)= (a\boxplus b)\wedge(a\boxplus c)$. {\bf\hfill Q.E.D.}


\section{Variants}\label{sec:4}

\begin{definition}\rm\label{def:4}
Let $M=(Q,\Sigma,\Gamma,\delta,B,I,T)$ be an $\mathcal{E}\mbox{NTM}$. We call $\delta$ classical if $\delta(p,a,q,b,D)=0$ or 1 $\forall p,q\in Q$, $\forall a,b\in\Gamma$ and $\forall D\in\{L,S,R\}$. Similarly, we call $I$ ($T$) classical if $I(p)=0$ or 1 ($T(p)=0$ or 1) $\forall p\in Q$. The subclass of all $\mathcal{E}$NTMs with a classical initial-state (terminal-state) function is denoted as NTM$_I(\mathcal{E},\Sigma)$ (NTM$_T(\mathcal{E},\Sigma)$). We define NTM$_{IT}(\mathcal{E},\Sigma)=\mbox{NTM}_I(\mathcal{E},\Sigma)\cap\mbox{NTM}_T(\mathcal{E},\Sigma)$.
\end{definition}

The following results show that any $\mathcal{E}$NTM can be simulated by an $\mathcal{E}$NTM with a classical initial-state function. That is, $\mathcal{E}$NTMs with classical initial states are as powerful as general $\mathcal{E}$NTMs.

\begin{lemma}\rm\label{thm:1}
For any $M\in\mbox{NTM}(\mathcal{E},\Sigma)$ there exists $M_I\in\mbox{NTM}_I(\mathcal{E},\Sigma)$ such that $|M|_d=|M_I|_d$ and $|M|_w=|M_I|_w$.
\end{lemma}
{\bf Proof:} Assuming $M=(Q,\Sigma,\Gamma,\delta,B,I,T)$, we construct $M_I=(Q_I,\Sigma,\Gamma,\delta_I,B,I_I, T_I)$, where $Q_I=Q\cup\{p_I\}$ and $p_I\notin Q$,
\begin{align*}
  &I_I(p_I)=0,\mbox{ and }I_I(q)=1,\forall q\in Q\\
  &T_I(p_I)=1,\mbox{ and }T_I(q)=T(q),\forall q\in Q\\
  &\delta_I(p,a,q,b,D)=\delta(p,a,q,b,D),\forall p,q\in Q\\
  &\delta_I(p_I,a,q,a,S)=I(q),\forall q\in Q
\end{align*}
and $\delta_I=1$ for the rest. In $M_I$ the new state $p_I$ is the unique initial state. It is straightforward to see that $|M|_d=|M_I|_d$. We can directly prove the width-first method.
\begin{align*}
|M_I|_w(s)=&\bigwedge_{n\ge1}\bigg[\bigwedge_{C_n}\bigg(\cdots\bigg(\bigwedge_{C_1}
\bigg(\bigwedge_{q_0\in Q_I}I(q_0)\boxplus\delta_I^\star(q_0s,C_1)\bigg)\boxplus\delta_I^\star(C_1,C_2)\bigg)
\cdots\bigg)\boxplus T_I(q_n)\bigg]\\
=&\bigwedge_{n\ge1}\bigg[\bigwedge_{C_n}\bigg(\cdots\bigg(\bigwedge_{C_1}
\bigg(I_I(p_I)\boxplus\delta_I^\star(p_Is,C_1)\bigg)\boxplus\delta_I^\star(C_1,C_2)\bigg)
\cdots\bigg)\boxplus T_I(q_n)\bigg]\\
=&\ldots\\
=&\bigwedge_{n\ge1}\bigg[\bigwedge_{C_n}\bigg(\cdots\bigg(\bigwedge_{q_1\in Q}
I(q_1)\boxplus\delta^\star(q_1s,C_2)\bigg)
\cdots\bigg)\boxplus T(q_n)\bigg]\\
=&|M|_w(s)
\end{align*}
{\bf\hfill Q.E.D.}

Symmetrically, any $\mathcal{E}$NTM can be simulated by an $\mathcal{E}$NTM with a classical terminal-state function.

\begin{lemma}\rm\label{thm:7}
For any $M\in\mbox{NTM}(\mathcal{E},\Sigma)$ there exists $M_T\in\mbox{NTM}_T(\mathcal{E},\Sigma)$ such that $|M|_d=|M_T|_d$.
\end{lemma}
{\bf Proof:} Let $M=(Q,\Sigma,\Gamma,\delta,B,I,T)$ and $M_T=(Q_T,\Sigma,\Gamma,\delta_T,B,I_T,T_T)$, where $Q_T=\{(p,T(p)):p\in Q\}$,
\begin{align*}
  I_T((p,T(p))=&I(p)\\
  \delta_T((p,T(p)),a,(q,T(q)),b,D)=&
  \begin{cases}
    \delta(p,a,q,b,D)\boxplus T(q),&\mbox{ if }T(q)<1\\
    \delta(p,a,q,b,D),&\mbox{ if }T(q)=1
  \end{cases}\\
  T_T((p,T(p)))=&
  \begin{cases}
    0&\mbox{ if }T(p)<1\\
    1&\mbox{ if }T(p)=1
  \end{cases}
\end{align*}
and $\delta_T=1$ for the rest.

By Remark \ref{rmk:1}, an $\mathcal{E}$NTM halts in two cases: (i) it reaches some state $p$ such that $T(p)<1$ or (ii) it reaches some configuration $\alpha pa\beta$ such that $T(p)=1$ and $\delta(p,a,q,b,D)=1$ $\forall q,b,D$.

Let $s\in\Sigma^+$ be an arbitrary input and let $M$ halt along the path $P=(C_0=p_Is,C_1,\cdots,C_n)$. Suppose $\vdash(C_{i-1},C_i)=(p_{i-1},a_{i-1},p_i,a_i,D_i)$, $i=1,\cdots,n$. Then there is a path $M_T$: $P_T=(\tilde C_0=(p_I,T(p_I))s,\tilde C_1,\cdots, \tilde C_n)$, where $\tilde C_i=\alpha(p,T(p))\beta$ if $C_i=\alpha p\beta$ and $\vdash(\tilde C_{i-1},\tilde C_i)=((p_{i-1},T(p_{i-1})),a_{i-1},(p_i,T(p_i)),a_i, D_i)$, $i=1,\cdots,n$. If $M$ halts in case (i), then $T(St(C_n))<1$. Obviously $M_T$ halts along $P_T$ and the $\mathcal{E}$-values of $P$ and $P_T$ are the same. If $M$ halts in case (ii), then $T(St(C_n))=1$ and $\delta^\star(C_n,C')=1$ for all $C'$. By the definition, $\delta_T((p,T(p)),a,(q,T(q)),b,D)\geq\delta(p,a,q,b,D)$, so $\delta^\star_T(\tilde C_n,\tilde C')=1$ for all $\tilde C'$ and $T_T(St(\tilde C))=1$. Then $M_T$ also halts along $P_T$ and the $\mathcal{E}$-values of $P$ and $P_T$ all equal 1.

Conversely, assume that $M_T$ halts along the path $P_T=(\tilde C_0=(p,T(p))s, \tilde C_1,\cdots,\tilde C_n)$. Suppose that $\vdash(\tilde C_{i-1},\tilde C_i)=((p_{i-1},T(p_{i-1})),a_{i-1},(p_i,T(p_i)),a_i,D_i)$, $i=1,\cdots,n$. Then there is a path $P=(C_0=ps,C_1,\cdots,C_n)$ where $C_i=\alpha p\beta$ if $\tilde C_i=\alpha(p,T(p))\beta$. If $M_T$ halts along $P_T$ in case (i), i.e. $T_T(p_n,T(p_n))=0$, then $M$ halts along $P$ since $T(p_n)<1$ by definition and the $\mathcal{E}$-values of $P$ and $P_T$ are the same.

If $M_T$ halts along $P_T$ in case (ii), then $\delta_T((p_n,T(p_n)),a_n,(q,T(q)),b,D)=1$ for all $q,b,D$ and $T_T((p_n,T(p_n)))=1$. First, if $\delta_T((p_n,T(p_n)),a_n,(q,T(q)),b, D)=\delta(p_n,a_n,q,b,D)\boxplus T(q)=1$ for some $q,b,D$, we have $T(q)<1$ by definition. Then $M$ halts along $P'=(C_0,\cdots,C_n,C_{n+1})$, where $\vdash(C_n,C_{n+1})=(p_n,a_n,q,b,D)$ and the $\mathcal{E}$-value of $P'$ is 1. Otherwise, if $\delta_T((p_n,T(p_n)),a_n,(q,T(q)),b,D)=\delta(p_n,a_n,q,\\b,D)=1$ for all $q,b,D$, then $M$ halts along $P$ and the $\mathcal{E}$-value of $P$ is 1.

Therefore, we conclude that if $M$ halts along some path, then $M_T$ also halts along the ``mirror'' path with the same $\mathcal{E}$-value and vice versa. {\bf\hfill Q.E.D.}

Combining Lemmas \ref{thm:1} and \ref{thm:7}, we know that the non-classical parts of $\mathcal{E}$NTMs can exist only in the transition processes.

\begin{corollary}\rm\label{thm:5}
For any $M\in\mbox{NTM}(\mathcal{E},\Sigma)$ there exists $M_{IT}\in\mbox{NTM}_{IT}(\mathcal{E},\Sigma)$ such that $|M_{IT}|_d=|M|_d$.
\end{corollary}

Therefore, from now on we can denote an $\mathcal{E}$NTM by $M=(Q,\Sigma,\Gamma,\delta,B,p_I,T)$ if needed.

\begin{definition}\rm
A path $(C_0,\cdots,C_n)$ is effective if $\delta^\star(C_{i-1},C_i)\neq 1$ for $i=1,\cdots,n$. On an effective path, each $\delta^\star(C_{i-1},C_i)=\delta(St(C_{i-1}),a,St(C_i),b,D)$ for some $a,b\in\Gamma$ and $D\in\{L,S,R\}$.
\end{definition}

\begin{definition}\rm
Let $M=(Q,\Sigma,\Gamma,\delta,B,p_I,T)$ be an $\mathcal{E}$NTM. For any $s\in\Sigma^+$, we define $ID_M(s,1)=\{C\in ID(M):(p_Is,C)\mbox{ as an effective path}\}$ and $ID_M(s,n+1)=\{C\in ID(M):(C',C)\hbox{ as an effective path for}\\\hbox{some }C'\in ID_M(s,n)\}$, $n=1,2,\cdots$. Let $ID_M(s)=\bigcup_n ID_M(s,n)$ comprise all the IDs achievable from $p_Is$. We omit the subscript $M$ if no confusion is possible.
\end{definition}

From the definition above, Equation (\ref{eqn:1}) can be simplified to
\begin{align}\label{eqn:3}
|M|_d(s)=&\bigwedge_{n\ge1}\bigwedge_{C_i\in ID(s,n)} \delta^\star(p_Is,C_1)\boxplus\cdots \boxplus\delta^\star(C_{n-1},C_n) \boxplus T(St(C_n))
\end{align}
and Equation (\ref{eqn:2}) can be simplified to
\begin{equation}\label{eqn:4}
\begin{split}
|M|_w(s)=&\bigwedge_{n\ge1}\Bigg[\bigwedge_{C_n\in ID(s,n)}\Bigg(\cdots\Bigg(\bigwedge_{C_2\in ID(s,2)} \Bigg(\bigwedge_{C_1\in ID(s,1)}\delta^\star(p_Is,C_1)\boxplus\delta^\star(C_1,C_2)\Bigg)\\
&\boxplus\delta^\star(C_2,C_3)\Bigg)\cdots\Bigg)\boxplus T(St(C_n))\Bigg].
\end{split}
\end{equation}
We denote the range of a map $f$ by $R(f)$. For an $\mathcal{E}$NTM $M=(Q,\Sigma,\Gamma,\delta,B,I, T)$, let $R_M=R(I)\cup R(\delta)\cup R(T)$. We assume that $R_M=\{x_1,x_2,\cdots,x_k\}$ since it is finite. Thus, the value of path $P$ is $e(P)=v_1x_1\boxplus v_2x_2\boxplus\cdots\boxplus v_lx_k$, or simply represented by a $k$-vector $v(P)=(v_1,\cdots,v_k)$. Two $k$-vectors $(v_1,\cdots,v_k)$ and $(v'_1,\cdots,v'_k)$ are called compatible if $v_i\leq v'_i$ for all $i$, denoted by $(v_1,\cdots,v_k)\leq (v'_1,\cdots,v'_k)$. Obviously if $v(P_1)\leq v(P_2)$ then $e(P_1)\leq e(P_2)$. That is, in this case $P_2$ can be omitted from the calculus. A set of $k$-vectors is called independent if and only if all elements are not compatible with each other. In fact, Proposition 2 in \cite{W02} showed that any independent set of such $k$-vectors is finite. Thus, there are finite $\wedge$ operations in Equations (\ref{eqn:3}) and (\ref{eqn:4}).

Next we show that under some finiteness condition, each $\mathcal{E}$NTM can be simulated by some $\mathcal{E}$NTM with classical transitions.

\begin{theorem}\label{thm:14}\rm
Let $M$ be an $\mathcal{E}$NTM and let $S_M$ denote the subalgebra generated by $R_M$. If $S_M$ is finite, there exists an $\mathcal{E}$NTM $\bar{M}$ with classical transitions such that $|M|_w=|\bar{M}|_w$.
\end{theorem}
{\bf Proof.} Assume that $M=(Q,\Sigma,\Gamma,\delta,B,p_I,T)$. We construct $\bar{M}=(S^Q_M,\Sigma,\Gamma,\bar{\delta},B,\bar p_I,\bar{T})$ as follows:
\begin{align*}
\bar p_I(q)=&
\begin{cases}
0,&\mbox{if }q=p_I\\
1,&\mbox{otherwise}
\end{cases}\\
\bar{T}(X)=&\wedge_{p\in Q}X(p)\boxplus T(p)
\end{align*}
for any $a,b\in\Gamma$, $X\in S^Q_M$ and $D\in\{L,S,R\}$, $\bar{\delta}(X,a,Y,b,D)=0$, where $Y(q)=\wedge_{p\in Q}X(p)\boxplus\delta(p,a,q,b,D)\in S^Q_M$ and $\bar\delta=1$ for the rest. Here $\bar\delta$ can be treated as a classical transition function $S^Q_M\times\Gamma\longrightarrow 2^{S^Q_M\times\Gamma\times\{L,S,R\}}$.

We only need to consider effective paths $(Is,\bar C_1,\cdots,\bar C_n)$. For each effective path there exists a unique set $\{a_i,b_i,D_i\}^n_{i=1}$ satisfying $\bar{\delta}^\star(Is,\bar C_1)= \bar{\delta}(I,a_1,St(\bar C_1),b_1,D_1)$ and $\bar{\delta}^\star(\bar C_{i-1},\bar C_i)= \bar{\delta}(St(\bar C_{i-1}),a_i,St(\bar C_i),\\b_i,D_i)$ for $i=2,\cdots,n$. Thus, for any $s\in\Sigma^+$,
\begin{align*}
|\bar M|_w(s)=&\bigwedge_{n\ge1}\bigg[\bigwedge_{\bar C_n}\bigg(\cdots\bigg(\bigwedge_{\bar C_2}
\bigg(\bigwedge_{\bar C_1}\bar\delta^\star(\bar p_Is,\bar C_1)
\boxplus\bar\delta^\star(\bar C_1,\bar C_2)\bigg)\boxplus\bar\delta^\star(\bar C_2,\bar C_3)\bigg)
\cdots\bigg)\boxplus \bar T(St(\bar C_n))\bigg]\\
=&\bigwedge_{n\ge1}\bigwedge_{\bar C_n\in ID_{\bar M}(s,n)}\bar T(St(\bar C_n))
=\bigwedge_{n\ge1}\bigwedge_{\bar C_n}\bigwedge_{p_n\in Q}St(\bar C_n)(p_n)\boxplus T(p_n)\\
=&\bigwedge_{n\ge1}\bigwedge_{\bar C_n,p_n}\bigg(\bigwedge_{\bar C_{n-1}\in ID_M(s,n-1)}
\bigwedge_{p_{n-1}}St(\bar C_{n-1})(p_{n-1})\boxplus\delta(p_{n-1},a_n,p_n,b_n,D_n)\bigg)\boxplus T(p_n)\\
=&\bigwedge_{n\ge1}\bigg[\bigwedge_{\bar C_n,p_n}\bigg(\bigwedge_{\bar C_{n-1},p_{n-1}}\bigg(\cdots
\bigg(\bigwedge_{\bar C_1,p_1}St(\bar C_1)(p_1)\boxplus\delta(p_1,a_2,p_2,b_2,D_2)\bigg)\\
&\cdots\bigg)\boxplus\delta(p_{n-1},a_n,p_n,b_n,D_n)\bigg)\boxplus T(p_n)\bigg]\\
=&\bigwedge_{n\ge1}\bigg[\bigwedge_{\bar C_n,p_n}\bigg(\bigwedge_{\bar C_{n-1},p_{n-1}}\bigg(\cdots
\bigg(\bigwedge_{\bar C_1,p_1}\delta(p_I,a_1,p_1,b_1,D_1)\boxplus\delta(p_1,a_2,p_2,b_2,D_2)\bigg)\\
&\cdots\bigg)\boxplus\delta(p_{n-1},a_n,p_n,b_n,D_n)\bigg)\boxplus T(p_n)\bigg]\\
=&\bigwedge_{n\ge1}\bigg[\bigwedge_{\bar C_n,p_n}\bigg(\bigwedge_{\bar C_{n-1},p_{n-1}}\bigg(\cdots
\bigg(\bigwedge_{p_1,b_1,D_1}\delta(p_I,a_1,p_1,b_1,D_1)\boxplus\delta(p_1,a_2,p_2,b_2,D_2)\bigg)\\
&\cdots\bigg)\boxplus\delta(p_{n-1},a_n,p_n,b_n,D_n)\bigg)\boxplus T(p_n)\bigg]\\
=&\bigwedge_{n\ge1}\bigg[\bigwedge_{\bar C_n,p_n}\bigg(\bigwedge_{\bar C_{n-1},p_{n-1}}\bigg(\cdots
\bigg(\bigwedge_{C_1\in ID_M(s,1)}\delta(p_I,a_1,St(C_1),b_1,D_1)\boxplus\delta(St(C_1),a_2,p_2,b_2,D_2)\bigg)\\
&\cdots\bigg)\boxplus\delta(p_{n-1},a_n,p_n,b_n,D_n)\bigg)\boxplus T(p_n)\bigg]\\
=&\bigwedge_{n\ge1}\bigg[\bigwedge_{C_n\in ID_M(s,n)}\bigg(\cdots\bigg(\bigwedge_{C_1\in ID_M(s,1)}
\delta^\star(p_Is,C_1)\boxplus\delta^\star(C_1,C_2)\bigg)
\cdots\bigg)\boxplus T(St(C_n))\bigg]\\
=&|M|_w(s).\hspace{12cm}{\bf Q.E.D.}
\end{align*}

\begin{definition}\rm\cite{DP00}
A QMV algebra is said to be locally finite iff $\forall a\in\mathcal{E}$ s.t. $a\neq 0$ $\exists n\in\mathbb{N}$ s.t. $n\cdot a=1$.
\end{definition}

Let $M$ be an $\mathcal{E}$NTM. Let $R_M^\boxplus=\{a_1\boxplus a_2\boxplus\cdots\boxplus a_n: a_i\in R_M,n\in\mathbb{N}\}\cup\{0\}$. It is straightforward to prove that if $\mathcal{E}$ is locally finite, then $R_M^\boxplus$ is also finite. In the following we can simulate any $\mathcal{E}$NTM with some $\mathcal{E}$NTM with classical transitions.

After Corollary \ref{thm:5} the question arises as to whether the transitions of an $\mathcal{E}$NTM can be classical without losing power. The next lemma shows that this can be obtained under a certain finite condition.

\begin{lemma}\rm\label{thm:2}
Let $M$ be an $\mathcal{E}$NTM. If $\mathcal{E}$ is locally finite, there exists some $\mathcal{E}$NTM $M^c$ with classical transitions that accepts the same $\mathcal{E}$-valued language.
\end{lemma}
{\bf Proof. } Let $M=(Q,\Sigma,\Gamma,\delta,B,I,T)$ and $M^c=(Q^c,\Sigma^c,\Gamma^c,\delta^c,B,I^c,T^c)$. We assume that $||Q\times\Gamma\times Q\times\Gamma\times\{L,S,R\}||=N$ and we number all possible transitions $(p,a,q,b,D)$ from 1 to $N$.

The state set $Q^c=Q\cup\{q_x^{(i,j)}:q\in Q,x\in R_M^\boxplus,i=1,\cdots,N,j=0,\cdots,4\}\cup\{q_x^{(f)}: x\in R_M^\boxplus\}$ is finite since $R_M^\boxplus$ is finite. The input alphabet is $\Sigma^c=\Sigma\times\{0\}$, where $0$ is the least element of $\mathcal{E}$. The tape alphabet $\Gamma^c=\Sigma\times R_M^\boxplus\cup\{B\}$ is finite for finite $R_M^\boxplus$. The initial function is $I^c|_Q=I$ and $I^c|_{Q^c-Q}=1$ for the rest.

For each $\delta(p,a,q,b,D)=y$, suppose the index of $(p,a,q,b,D)$ is $i$. We define the following classical transitions:
\begin{align}
\delta^c(p,(a,x),q_{x\boxplus y}^{(i,0)},(b,x\boxplus y),S)=&0\label{eqn:6}\\
\delta^c(q_{x\boxplus y}^{(i,0)},(b,x\boxplus y),q_{x\boxplus y}^{(i,1)},(b,x\boxplus y),L)=&0\label{eqn:7}\\
\delta^c(q_{x\boxplus y}^{(i,1)},(c,z),q_{x\boxplus y}^{(i,2)},(c,x\boxplus y),R)=&0,
\forall c\in\Gamma,z\in R^\boxplus_M\label{eqn:8}\\
\delta^c(q_{x\boxplus y}^{(i,2)},(b,x\boxplus y),q_{x\boxplus y}^{(i,3)},(b,x\boxplus y),R)=&0\label{eqn:9}\\
\delta^c(q_{x\boxplus y}^{(i,3)},(c,z),q_{x\boxplus y}^{(i,4)},(c,x\boxplus y),L)=&0,
\forall c\in\Gamma,z\in R^\boxplus_M\label{eqn:10}\\
\delta^c(q_{x\boxplus y}^{(i,4)},(b,x\boxplus y),q,(b,x\boxplus y),D)=&0\label{eqn:11}\\
\delta^c(q,(c,z),q_z^{(f)},(c,z),S)=&0,\forall c\in\Gamma,z\in R^\boxplus_M\label{eqn:12}
\end{align}
and $\delta^c=1$ for the rest. Finally, $T^c(q_x^{(f)})=x\boxplus T(q)$ and $T^c(p)=1$ for the rest. Assume that $M$ can transform from ID $\alpha pa\beta$ to $\alpha qb\beta$ through the transition $\delta(p,a,q,b,D)=y$. Let $\bar\alpha=\bar\alpha'(c_1,z_1)$ and $\bar\beta=(c_2,z_2)\bar\beta'$; then $M^c$ must run as follows:
\begin{align*}
&\bar\alpha p(a,x)\bar\beta\stackrel{(\ref{eqn:6})}{\longrightarrow}
\bar\alpha q^{(i,0)}_{x\boxplus y}(b,x\boxplus y)\bar\alpha\stackrel{(\ref{eqn:7})}{\longrightarrow}
\bar\alpha'q^{(i,1)}_{x\boxplus y}(c_1,z_1)(b,x\boxplus y)\bar\beta\stackrel{(\ref{eqn:8})}{\longrightarrow}
\bar\alpha'(c_1,x\boxplus y)q^{(i,2)}_{x\boxplus y}(b,x\boxplus y)\bar\beta\stackrel{(\ref{eqn:9})}{\longrightarrow}\\
&\bar\alpha'(c_1,x\boxplus y)(b,x\boxplus y)q^{(i,3)}_{x\boxplus y}(c_2,z_2)\bar\beta'
\stackrel{(\ref{eqn:10})}{\longrightarrow}
\bar\alpha'(c_1,x\boxplus y)q^{(i,4)}_{x\boxplus y}(b,x\boxplus y)(c_2,x\boxplus y)\bar\beta'
\stackrel{(\ref{eqn:11})}{\longrightarrow}\\
&\begin{cases}
\bar\alpha'q(c_1,x\boxplus y)(b,x\boxplus y)(c_2,x\boxplus y)\bar\beta',&\mbox{if }D=L\\
\bar\alpha'(c_1,x\boxplus y)q(b,x\boxplus y)(c_2,x\boxplus y)\bar\beta',&\mbox{if }D=S\\
\bar\alpha'(c_1,x\boxplus y)(b,x\boxplus y)q(c_2,x\boxplus y)\bar\beta',&\mbox{if }D=R.
\end{cases}
\end{align*}

Since $M^c$ is non-deterministic, transition (\ref{eqn:12}) would take the machine into state $q^{(f)}_{x\boxplus y}$ and then it must halt. To see this, if $T^c(q^{(f)}_{x\boxplus y})=x\boxplus y\boxplus T(q)<1$, then $M^c$ halts. Otherwise the machine is in state $q^{(f)}_{x\boxplus y}$ and then the $\mathcal{E}$-values of all the next possible transitions are $1$, so $M^c$ must halt. Therefore, we can see that $M^c$ turns into $\tilde\alpha q(b,x\boxplus y)\tilde\beta$ from $\bar\alpha p(a,x)\bar\beta$ through transitions (\ref{eqn:6})--(\ref{eqn:11}).

Now suppose the input is $s$ and there is an effective path for $M$:
\begin{align*}
&I(p_0)\boxplus\delta^\star(p_0s,C_1)\boxplus\cdots\boxplus\delta^\star(C_{n-1},C_n)\boxplus T(p_n)\\
=&I(p_0)\boxplus\delta(p_0,a_1,p_1,b_1,D_1)\boxplus\cdots\delta(p_{n-1},a_n,p_n,b_n,D_n)\boxplus T(p_n).
\end{align*}
According to the above discussion, there is an effective path for $M^c$:
\begin{align*}
&I(p_0)\boxplus\delta^{c\star}(p_0s\times\{0\},\bar C_1)\boxplus\cdots\boxplus
\delta^{c\star}(\bar C_{n-1},\bar C_n)\boxplus T^c(St(\bar C_n))\\
=&I(p_0)\boxplus 0 \boxplus\cdots\boxplus 0 \boxplus T^c(p^{(f)}_x)\\
=&I(p_0)\boxplus\delta(p_0,a_1,p_1,b_1,D_1)\boxplus\cdots\delta(p_{n-1},a_n,p_n,b_n,D_n)\boxplus T(p_n),
\end{align*}
where
$x=\delta(p_0,a_1,p_1,b_1,D_1)\boxplus\cdots\boxplus\delta(p_{n-1},a_n,p_n,b_n,D_n)\boxplus T(p_n)$. The $\mathcal{E}$-values of these two paths are the same.

Conversely, any $M^c$ input must be in the form $s\times\{0\}$, where $s\in\Sigma^+$, so each effective path for $M^c$ can be simulated by some path of $M$.{\bf\hfill Q.E.D}

Using the same construction as in Lemma \ref{thm:2}, we can show that if $M$ is deterministic, then $M^c$ can also be deterministic.

\begin{corollary}\label{thm:6}\rm
Let $M$ be an $\mathcal{E}$DTM. When $\mathcal{E}$ is locally finite, there exists some $\mathcal{E}$DTM $M^c$ with classical transitions that accepts the same $\mathcal{E}$-valued language.
\end{corollary}

In fact we can assume that $M$ in Lemma \ref{thm:2} has a single initial state by Lemma \ref{thm:1}, and therefore $M^c$ has a single initial state and a classical transition function.

In classical computation theory, deterministic Turing machines are equivalent to non-deterministic Turing machines, that is, they can recognize the same languages. However, this property does not hold for fuzzy non-deterministic Turing machines \cite{W02,L08}. Fuzzy non-deterministic Turing machines are more powerful than fuzzy deterministic Turing machines. Similarly, we show that $\mathcal{E}$NTMs are also more powerful than $\mathcal{E}$DTMs.

Let $\mathcal{E}$ be locally finite. By Lemmas 4.1 and 4.5, we can assume that $M$ is an $\mathcal{E}$DTM with classical transitions and a single initial state. Then we can construct a classical Turing machine with two tapes to compute the $\mathcal{E}$-valued language $|M|_d$. For any input $s$, in the first tape, $M'$ simulates $M$ according to the transition function of $M$. Since $M$ is deterministic, the $\mathcal{E}$ value of each step can be recorded in the second tape. Obviously, $M'$ halts iff $M$ halts. When $M'$ halts, the final result for the second tape is just $|M|_d(s)$.

From the above discussion, we can conclude that there exists $\mathcal{E}$DTM that can be simulated by a classical Turing machine. However, in the following example we find that for some $\mathcal{E}$NTM, there is no classical Turing machine that can simulate it.

\begin{example}\label{exp:1}\rm
Let $L_u$ be the standard \emph{universal language} in classical computation theory and let $M_u=(Q_u,\Sigma,\Gamma,\delta_u,B,p_I,Q_T)$ be the universal Turing machine accepting $L_u$. We construct an $\mathcal{E}$NTM $M=(Q,\Sigma,\Gamma,\delta,B,q_I,T)$ such that, for any given $0<x<1$,
\begin{itemize}
\item $Q=Q_u\cup\{q_I,q_T\}$, where $q_I,q_T\notin Q_u$.
\item $\delta(q_I,a,p_I,a,S)=\delta(q_I,a,q_T,a,S)=0$ $\forall a\in\Sigma$.
\item $\delta(p,a,q,b,D)=0$ if and only $(q,b,D)\in\delta_u(p,a)$, and $\delta=1$ for the others.
\item $T(p)=0$ for $p\in Q_T$, and $T=1$ for the others.
\end{itemize}
Obviously $M$ is an $\mathcal{E}$NTM and its language is $|M|_d(s)=0$ $\forall s\in L_u$ and $|M|_d(s)=x$ $\forall s\notin L_u$. If there exists some $\mathcal{E}$DTM $M'$ simulating $M$, then the classic language $\{s\in\Sigma^*:|M'|_d(s)=x\}=\Sigma^*-L_u$ must be recursively enumerable, which contradicts the fact that $L_u$ is undecidable.
\end{example}
As a result, we obtain the following theorem.

\begin{theorem}\rm
$\mathcal{E}$NTMs are not equivalent to $\mathcal{E}$DTMs and $\mathcal{E}$NTMs have more computational power than $\mathcal{E}$DTMs.
\end{theorem}


\vspace{-.6em}  

\section{Conclusion}\label{sec:5}
To set up a quantum computation theory for characterizing open quantum systems, we continue to discuss Turing machines based on unsharp quantum logic. By reexamining some properties of classical Turing machines, we found that some important properties are different from those of classical Turing machines, such as the relation between $\mathcal{E}$NTMs and $\mathcal{E}$DTMs. We also found that some $\mathcal{E}$NTMs with some classical characters have the same power as general $\mathcal{E}$NTMs. The phrase structure grammar, the universality of the Turing machines, the multitape case and the closure properties of unsharp Turing machines will be presented elsewhere.

\vspace{-.6em}  


\nocite{*}
\bibliographystyle{eptcs}
\bibliography{Turing_machines_based_on_unsharp_quantum_logic}
\end{document}